\documentclass[ aps, reprint, amsmath, amssymb, floatfix ]{revtex4-1}
\usepackage{graphicx}
\usepackage{dcolumn}
\usepackage{bm}
\usepackage{babel}

\begin{document}


\title{ On the status of channeling radiation and laser based radiation sources  }
\author{Khokonov M.Kh.}
\altaffiliation{Kabardino-Balkarian State University, Nalchik, Russian Federation.}
 \email{Electronic address: khokon6@mail.ru}     
\date{\today}

\begin{abstract}
A unified description of channeling radiation (CR) in oriented crystals (OC) and radiation in the 
field of an intense plane wave (laser radiation sources -- LRS) is proposed in terms of two Lorentz-invariant parameters. 
The case of planar positron channeling (quasi-channeling) and linearly  polarized laser wave is considered in detail up to TeV energies. 
The crucial difference between LRS and CR is that in the former case both invariants are independent, while in channeling they are linearly related to each other. This leads to a strong limitation on the range of possible values of these invariants in OC. On the other hand, OC make it possible to study QED processes 
in strong non-uniform external fields.
\end{abstract}

\maketitle

\section{Introduction}

Channeling radiation (CR) in oriented crystals (OC) \cite{Beloshitsky1982, Ulrik2005} provides a promising source of intense gamma radiation, as well as  makes it possible to experimentally study such phenomena as the radiation reaction  
  \cite{Ulrik_PLB_2017, Ulrik_NC_2018, mkh_2019, Wistisen_2019, Wistisen_2020},  strong field effects \cite{Ulrik2005, 
uggerhoj_2020, uggerhoj_2021} and  
trident production \cite{trident_2010, trident_2022}.   
 Over the past two decades, interest in interaction of relativistic electrons with intense 
laser fields (referred to below as the laser radiation sources, LRS) has grown significantly \cite{Keitel_RMP2012, LUXE_2021} as an alternative method of  producing X-ray and gamma radiation \cite{mkh_2002, King_2020}, and as a tool   for studying the QED strong field effects \cite{Ritus1985, Gonoskov_RMP_2022, Piazza_PRD_2019, Ilderton_PRD_2019, Mironov_PRD_2019, 
Kostykov_Sci_2019, Keitel_PRL_2010, Ilderton_PRL_2019}. Therefore, it is of interest to establish the status of channeling and lasers in the context of the problems outlined.

Close similarity between CR  in crystals and lasers has been studied in Refs. \cite{mkh_2002, Carrigan1998}. During channeling, a relativistic electron interacts with a static electric field of atomic chains or planes, while for lasers, the interaction occurs with a field close to the field of a plane wave.
The appearance of higher harmonics in the radiation spectra arising in the field of an intense laser wave 
is due to the process of absorption of  several photons from  the laser field with  subsequent  emission  a photon whose energy is Doppler shifted toward the harder  frequency range.  
The same process happens in OC, where an electron  absorbs the virtual photons of the electrostatic atomic string (plane) potential of a crystal.


The external field is considered strong if the Schwinger's field parameter, $\chi$, exceeds unity, $\chi  =  e\hbar|F_{\mu\nu}p^\nu| /(m^3 c^4)$, where  $F_{\mu\nu}$ is the field electromagnetic tensor, $p^\nu$ is the 4-momentum of the electron, $e$ and $m$ are the charge and rest mass of the electron, $c$ is the velocity of light in vacuum. 
In oriented crystals, $\chi=\hbar F \gamma /(m^2c^3)$, where $\gamma=E/mc^2$, is the Lorenz-factor of the electron with energy $E$, $F=|\nabla U|$ is the force acting on the electron from the continuum potential, $U$, of atomic axes or planes of the crystal. In the case of axial channeling, $F\approx 2Ze^2/(da_F)$,  where $Z$ is the atomic number of the crystal,  $d$ is the distance between atoms in the atomic string, $a_F$ is the Thomas-Fermi screening parameter. For example, in silicon $\langle 110 \rangle$  crystal ($Z=14$), $F\approx 520$ eV/$\AA$,   and $\chi \approx 1$ for electron energies $E\approx$150 GeV.       
The strong field effects in radiation, like quantum recoil due to hard photon emission, 
are significant already at $\chi >0.1$. In what follows the Lorentz-factor corresponding to the initial energy of an electron (positron) before the  interaction with the external field will be denoted by $\gamma$, whereas the current time-dependent  Lorentz-factor will be $\gamma (t)$ (or, $\gamma (\delta)$, $\delta$ is the invariant phase of a plane wave).

In contrast with LRS channeling radiation occurs only at relativistic energies.  
Let an electron undergoing arbitrary extreme rela\-ti\-vis\-tic mo\-tion with $\gamma =(1-\beta^2)^{-1/2}  \gg 1$ 
en\-coun\-ter the external  field with a potential $|U|\ll E$, $\beta=v/c$, $v$ is the electron's velocity.  The angle of deviation of the electron
by the external field, $\theta_e$,  is then small enough such that the component of  electron's
velocity transverse to the velocity of the average rest frame (ARF) 
is non-relativistic, i.e. $\beta_\perp =v_\perp/c \approx \theta_e \ll 1$.
As is well known the peculiarities of the radiation spectra of relativistic electrons  depend on
the so-called ``non-dipole parameter'' $D=\beta_\perp 
\gamma \approx \theta_e \gamma$, $D$ is Lorentz-invariant, since the product, $\gamma \beta_\perp$, is invariant, \cite{Jackson} (chapter 11.4). 
If the deflection angle is larger than the characteristic radiation angle $\sim 1/\gamma$, then the condition
$D \gg 1$ is fulfilled, which means that only a small part of the
electron trajectory defines the emission spectrum which is described by  simple synchrotron like formulas (constant field approximation -- CFA). 
In the opposite limit, $D\ll 1$, the dipole approximation is valid, and the formulas for the emission spectrum are also substantially simplified.
 Note, that the ``longitudinal'' Lorentz-factor, corresponding to the velocity of the ARF, $v_0$, $\gamma_0=(1-\beta_0^2)^{-1/2}$,   is in connection with the total Lorenz-factor as $\gamma = \gamma_0 (1+D^2)^{1/2}$ (in the absence of the longitudinal oscillations, such that, $\beta^2=\beta_0^2+\beta_\perp^2$).   We will  assume that $\gamma \gg 1$ and $\gamma_0 \gg 1$.  In what follows, the angle brackets will denote averaging over the period, $T$, of the electron transverse oscillations, $\langle (...) \rangle =T^{-1} \int_0^T (...)dt$.  The precise definition of the invariant, $D$, in the present paper is,
$D=\langle \beta_\perp (t)^2 \gamma (t)^2 \rangle^{1/2}$, both for CR and LRS.

The intensity of a plane electromagnetic wave is characterized by the Lorentz-invariant parameter, 
$\nu_0=e{ \cal E}_0/(\sqrt{2} mc\omega_0)$, where $ {\cal E}_0$ is the amplitude of the electric field of the plane wave with frequency $\omega_0$.
As we will see below, the exact relation holds for LRS, $\nu_0 = D$. This equality underlies the comparison of CR  with LRS (see also Refs. \cite{Carrigan1998,  mkh_NIMB1998}).
We also assume that  $\gamma \gg D$, which provides a small-angle electron scattering by the external field,  $\theta_e \ll 1$.
 In channeling this condition is always satisfied, since charged particles are  incident on a single crystal with small angles to crystallographic directions and $\theta_e \sim  \theta_L=(2U_m/E)^{1/2}$, $\theta_L$ being the  Lindhard critical channeling angle \cite{Beloshitsky1982, Ulrik2005, Lindhard1965}, and  $U_m$ being the depth of the continuum potential well of atomic planes or strings.   
For LRS, $\theta_e \sim \nu_0/\gamma$, and  the condition, $\theta_e \ll 1$,  is violated for petawatt lasers at energies below  hundreds  MeV.

We will need one more invariant, which depends on the electron energy, $a=2\hbar \gamma \Omega_0 /(mc^2)$, where
$\Omega_0= 2\pi/T$, is the period of the transverse oscillations.  For LRS it can be written in the form, $a=2\hbar (k_0 \! \cdot \! p)/ (mc)^2$, 
where $k_0$ and $p$ are the 4-wave vector of the incoming plane wave and 4-momentum of the electron. 

\section{Equations of motion}

The trajectory of  ultrarelativistic particle interacting with a relatively weak external field, $|U|\ll E$ (and  $\theta_e \ll 1$), can be conveniently divided into transverse and longitudinal parts. 
We direct the longitudinal component along the $z$ axis, which in the case of channeling is parallel to the atomic chains (or planes) of the crystal, and for LRS it coincides with the direction of the initial electron (positron) velocity. 
The longitudinal velocity of the electron is then expressed in terms of its transverse velocity 
\begin{equation}
\beta_z (t) \approx 1 - (2\gamma^2)^{-1} - \mbox{\boldmath $\beta$}_\perp^2 (t)/2, 
\label{longitudinal_velocity}
\end{equation}
where
 $ c \mbox{\boldmath $\beta$}_\perp = d{\bf r}_\perp /dt$,  
${\bf r}_\perp$ is the transverse coordinate. 
The basic correction to Eq.(\ref{longitudinal_velocity}) is equal to, $-(U/E)\gamma^{-2}$, which  is negligible compared to the last two terms in Eq.(\ref{longitudinal_velocity}). 
The subsequent terms are of the order of, $\sim \beta_\perp^4$, $ (\beta_\perp /\gamma)^2$, $\gamma^{-4}$. 


The transverse motion in OC is non-relativistic and satisfies Newton's equations with a relativistic mass, 
$m\gamma d^2 {\bf r}_\perp / dt^2 = {\bf F}$, where  ${\bf F}=-\nabla U$. 
Conventionally, for channeling equation (\ref{longitudinal_velocity}) 
is considered as exact equation \cite{Beloshitsky1982}. 
For LRS  ${\bf F}$ represents the electromagnetic Lorentz force (see Supp.A). 
The trajectory, ${\bf r}(t)= ({\bf r}_\perp (t), z(t))$,  is expressed in terms of its transverse part, where the longitudinal coordinate, $z(t)$ is obtained by integration of Eq.(\ref{longitudinal_velocity}) over time.
In the absence of secondary effects, such as the  multiple scattering and radiation damping, 
the transverse energy in channeling, $E_\perp = {\bf p}_\perp^2/(2\gamma m) +U({\bf r}_\perp)$, 
is the integral of  motion, as is the longitudinal momentum,
$p_z=\gamma mc \beta_z $ \cite{Beloshitsky1982}. 
We do not consider the secondary factors.

The limits of applicability of equation (\ref{longitudinal_velocity}) 
are
\begin{equation}
\gamma \gg 1, \,\,\,\,\,\,\,\,\,\,\,\,\,\gamma \gg D. 
\label{inequalities}
\end{equation}
These relations summarize inequalities given in the Introduction. 

We will compare the planar channeling of positrons with electrons moving head-on the linearly polarized laser field. Positrons move in the continuum potential of the atomic planes, which potential is close to the  parabolic,   $U(x)=4U_m x^2/d_p^2$,  \cite{Beloshitsky1982, Lindhard1965}, where $x$ is the transverse coordinate, $d_p$ is the distance between atomic planes.
The transverse motion of positrons is one-dimensional and occurs along the $x$ axis. The origin of the coordinate system lies between the atomic planes along which the $z$ axis is directed. 
Radiation for planar channeling has been studied in detail in Refs. \cite{Beloshitsky_Kumakhov_1978, Zhevago1978, 
Kumakhov_Trikalinos_1980}, and in the recent papers \cite{Wistisen_oscillator_2018, Wistisen_planar_2019}. 

For LRS electrons are assumed to move along $z$ axis and interact head on with a plane wave which electric field,  
$\mbox{\boldmath ${\bf \cal E}$}=- \mbox{\boldmath ${\bf \cal E}$}_0 \sin \delta $, is directed parallel to $x$-axis, the invariant  phase here is, $\delta=\omega_0 (t +z/c)$.



%

Equations of motion, both for positron channeling and for LRS, in the external fields described above, within the limits of applicability of Eqs. (\ref{inequalities}) are
\begin{eqnarray} 
x (t) & = & x_m \, \sin (\Omega_0 t)  , \nonumber \\
z (t)&  = &  ct  \beta_0  - \frac{x_m^2 \Omega_0}{8c} \, \sin (2\Omega_0 t), 
\label{pl_las_eq_motion} \\
\beta_0 &=&  \langle \beta_z \rangle =
1 - (2\gamma^2)^{-1} -   \langle \beta_\perp^2 \rangle/2,  \nonumber
\end{eqnarray}
\vspace{3mm}
where $\langle \beta_\perp^2 \rangle=(x_m \Omega_0)^2/2c^2$,  $\beta_0$ is the velocity of the ARF. Different quantities in Eqs.    (\ref{pl_las_eq_motion}) for channeling and LRS are given in the Table 1.




\begingroup

\begin{center}
\begin{tabular}{c | c c}  
\multicolumn{3}{l}{Table 1. Correspondence between different quantities}\\
\multicolumn{3}{l}{in the field of lasers and planar positron channeling. }\\
\multicolumn{3}{l}{Here, $\gamma$ is the initial Lorentz-factor, for LRS  }\\
\multicolumn{3}{l}{$\gamma = \gamma_0 (1+\nu_0^2)^{1/2}$, $\Omega_0$ is the frequency  of transverse }\\
\multicolumn{3}{l}{oscillations, $\theta_L=(2U_m/E)^{1/2}$, $E=\gamma m c^2$}\\
\hline
& Laser & Channeling  \\      
 \hline
 & & \\
$x_m$  &  $\sqrt{2}$ \mbox{\large  $\frac{\nu_0 c}{\Omega_0 \gamma}$  } &  \mbox{ {\large  $\frac{d_p}{2}
\! \left( \frac {E_{\perp}}{U_m}  \right)^{\mbox{\tiny 1/2}}$} }   \\ 
 & & \\
$\Omega_0$ & $(1+\beta_0)\omega_0 $ & \mbox{\large  $\frac{2 c}{d_p}$  }$\!\!\theta_L$  \\    
 & & \\
$\chi$&  \mbox{\large  $\frac{e{\cal E}_0 \hbar \gamma }{m^2 c^3}$  }$\!\!(1+\beta_0)$  &
   \mbox{\large  $\frac{ \mbox{ \small $\mid \nabla U \mid$} }{m^2 c^3}$  }$\!\! \hbar\gamma$ \\  
 & & \\
$D$  & $\nu_0$  &   \mbox{\large  $\frac{(E_\perp E)^{1/2}}{m c^2}$  }    \\  
 & & \\
$a$  & \mbox{\large  $\frac{2\hbar \omega_0 }{mc^2}$  }$\!\!(1+\beta_0)\gamma$  & 
 \mbox{ {\large  $\frac{4\hbar }{mc d_p} \! \left( \frac {2U_m \gamma}{mc^2} \right)^{\mbox{\tiny 1/2}}$ }  }   \\  
\end{tabular}
\end{center}
\endgroup

The comparison of Eqs. (\ref{pl_las_eq_motion}) with exact classical equations of motion for LRS  is given in Supp.A.

\section{Radiation spectrum}  

We calculated the radiation spectra within the frame of the quasi-classical method of Baier and Katkov (BK) \cite{Baier1998} (see Supp.B), which applicability is consistent with conditions (\ref{inequalities}). The alternative formulation of the quasiclassical method is given in Ref. \cite{Lindhard_91}.  
As is shown in \cite{Artru2019} the BK method is accurate for LRS, whereas this may not be the case in free-to-bound transitions in OC \cite{Artru2019, Artru2015} as well as for hard photon radiation, $u \rightarrow 1$.


For a quasiperiodic motion, ${\bf r}_\perp (t)={\bf r}_\perp (t+T)$, given by Eqs. (\ref{pl_las_eq_motion}), 
the photon number spectrum emitted per unit phase of the transverse motion, $\psi=\Omega_0 t$,  both for CR and LRS, as a function of invariant parameters $D$ and $a$ only, has the form 
(see also Supp.B)
\begin{equation}
\! \! \! \frac{ d^2 N_\gamma (u, a, D)}{N_0 \, dud\psi d\varphi } = \frac{3}{\pi a} \!
\sum_{k=k_m}^{\infty} \!\! 
\left[ \left( 1 \! + \! \frac{uu'}{2} \right) g_{k} \! + \! \frac{uu'}{4D^2} j_{zk}^2 
 \right] \! ,
\label{photon_spectrum}
\end{equation}
\vspace{-5mm}
\begin{equation}
g_{k}=j_{xk}^2 + \frac{\eta_k^2}{2D^2}\, j_{zk}^2 - \sqrt{2}
\,\frac{\eta_k}{D}\,j_{xk}\,j_{zk}\,  \cos \varphi \, ,
\label{class_contribution_to_photon_spectrum}
\end{equation}
where $u=\hbar \omega/E$ is the photon energy measured in the units of electron's
initial energy, $u'=u/(1-u)$, $\eta_k^2=ak/u'-\nu_0^2-1$, $\eta_k=\gamma \theta_k$, $\theta_k$ is the polar angle of radiation for $k$-th harmonic, $\varphi$ is the azimuth angle of radiation, $k_m =1+E(u'(1+D^2)/a)$, $E(\eta)$ is the integer part of $\eta$. 
$N_0= dN_\gamma^{class}/d\psi=(2/3)\alpha D^2 $ is the number of emitted quanta in the classical limit of Thomson back-scattering, $\alpha = 1/137$. 
For fixed harmonic number, $k$, the emitted photon energy changes within
the interval  
\begin{equation}
0<u<u_{mk}=ak/(1+D^2+ak). 
\label{u_mk}
\end{equation}
Exact formula for frequency emitted by an electron for LRS beyond the conditions (\ref{inequalities})
is given in  \cite{Ritus1985, Harvey2009}.
 
Fourier components of the electron current $j_{xk}$ and $j_{zk}$ are expressed through the Bessel functions $J_m (x)$:  
$$
j_{xk} =  B^{-1}\,\sum_{m=-\infty}^{\infty} \,(k+2m)\,
J_m (A)\,J_{k+2m}(B) \, , \nonumber
$$
\vspace{-10mm}
\begin{equation}
\label{Fourier_components_of_currents}
\end{equation}
\vspace{-11mm}
$$
 j_{zk}  =  \sum_{m=-\infty}^{\infty} \,  
J_m (A)\,J_{k+2m}(B) \, , \nonumber
$$
where $A=(2a)^{-1} D^2 u'$, $B=2\sqrt{2} \,a^{-1} D
u' \eta_k \cos  \varphi $.  
Expressions like (\ref{Fourier_components_of_currents}) are well known in the  theory of 
LRS \cite{Ritus1964a}, planar positron channeling \cite{Beloshitsky1982, Baier1980,  Planar_positrons_1981} and undulator radiation \cite{Walker_1998}. Formulas (\ref{photon_spectrum}) -- (\ref{Fourier_components_of_currents}) are greatly simplified in two limiting cases: $D \ll 1$ (dipole approximation) and $D \gg 1$ (CFA).

In the limit $D\ll 1$  Eqs. (\ref{photon_spectrum}) -- (\ref{Fourier_components_of_currents}) give a quantum dipole spectrum
\begin{equation}
\frac{d^2 N_\gamma}{N_0 \, dud\psi} = \frac{3}{2a}\,\left(1-
2\frac{u'}{a}+ 2\frac{u'^2}{a^2}+\frac{uu'}{2} \right) \, ,
\label{quantum_dipole_spectrum}
\end{equation}
where $0<u<a/(1+a)$. The shape of the dipole spectrum depends only on one invariant $a$. 

In the opposite limit of CFA, $D\gg 1$,  harmonics with $k \gg 1$  play a decisive role.  In this case the spectrum has the form of the well-known quantum synchrotron formula  \cite{Klepikov1954}
\begin{equation}
\frac{d^2 N_\gamma^{(CFA)}}{N_0 dud\psi} = \frac{\sqrt{3}}{\pi a
D^2}
\label{CFA_spectrum}
\end{equation}
\vspace{-5mm}
$$ 
\times \, \Bigl[ (2+uu') K_{2/3}(\xi) -
 \int_{\xi}^\infty \,K_{1/3}(\eta)d\eta \Bigr] \, ,  
$$
where  $\xi=2u'/(3\chi)$. Here, the field parameter $\chi$ is expressed through invariants $a$ and $D$:
$\chi=a D \sin \Omega_0 t /\sqrt{2}$. Eq.(\ref{CFA_spectrum}) should be averaged over the period of the electron transverse motion (i.e. over $\Omega_0 t$).
Convenient representation of Eq. (\ref{CFA_spectrum}) without special functions is given in \cite{Khokonov1997, Khokonov2004}.    The spin contribution 
(i.e. electron (positron) helicity-flip contribution)  
in Eqs. (\ref{photon_spectrum}), (\ref{quantum_dipole_spectrum}) and (\ref{CFA_spectrum}) is  due to the terms containing the product $uu'$. 

\vspace{-5mm}

\section{The $D \div a$ diagram }

Despite the similarity of the CR and LRS, outlined in the previous sections, there is a significant difference between them, associated with the different behavior of invariants $a$ and $D$ as  functions of the electron (positron) energy. For LRS $D$ and $a$ are independent parameters, $D$ does not depend on energy and $a \sim \gamma$.  
For channeling, both $D$ and $a$, demonstrate the same type of energy dependence, proportional to $\gamma^{1/2}$, because in this case, $\Omega_0 \sim \gamma^{-1/2}$. It means that in channeling $D$ and $a$ are not independent and represent a line in the ($D \div a$) space.  Further, we will also be interested in the axial channeling of electrons, which we consider approximately. In this case we will take  $U_m \simeq 2Ze^2/d $,
$\theta_L=(4Ze^2/dE)^{1/2}$, $\Omega_0 \approx c\theta_L/a_F$. 
By means of these relations and formulas of the Section 2 we obtain the connection between $a$ and $D$ for axial and planar channeling 
\begin{eqnarray}
a & = &   (4 \lambdabar_c /d_p)(2U_m/E_\perp)^{1/2} D \,, \, \mbox{planar positrons},
\label{a_D_connection_planar}
 \\ 
a & \approx & (2\lambdabar_c /a_F) D  \,, \,\,\,\,\,\,\,\,\,\,\, \mbox{axial electrons},
\label{a_D_connection_axial}
\end{eqnarray}
where $\lambdabar_c = \hbar/mc$ is the Compton wave length. The planar Eq.(\ref{a_D_connection_planar}) is exact, whereas the axial Eq.(\ref{a_D_connection_axial}) is approximate. Explanations for axial case are given in Supp.C.  For channeling, only those values of $a$ and $D$ are possible  that satisfy Eqs.(\ref{a_D_connection_planar}) and (\ref{a_D_connection_axial}), while there are no restrictions for LRS.

\begin{figure}
\centering
\includegraphics[bb=80 0 696 750, scale=0.52]{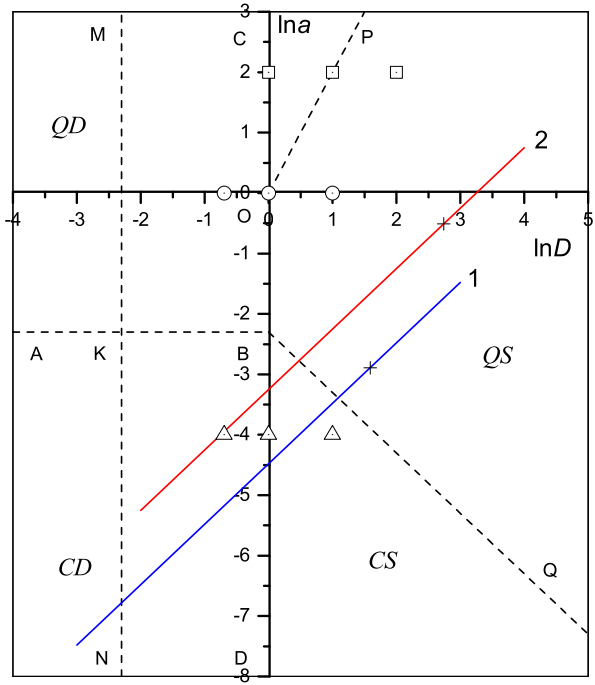}
\centering
\vspace{-55mm}
\caption{
The $D \div a$ diagram. Different regions are explained in the text. Lines 1 and 2 correspond to planar positron, Eq.(\ref{a_D_connection_planar}), and axial electron, Eq.(\ref{a_D_connection_axial}) channeling, respectively. 
}
\label{fig_classification}
\end{figure}

The above results can be conveniently illustrated using a diagram in the $a$ versus $D$ plane on the logarithmic scale similar to that done in  Ref.\cite{Khokonov2005} for LRS.  A diagram presented in the Fig.\ref{fig_classification} classifies CR and LRS in terms of invariants $X \equiv \ln D$ (abscissa) and $Y\equiv \ln a$ (ordinate), so that the origin of the coordinate system corresponds to the $D$ and $a$ values of unity. In this coordinate system, the region of strong fields is to the right of the $Y$- axis, while the region where quantum effects in the radiation spectrum occur lies in the vicinity of the $X$- axis and above this axis.
The region to the left of the dashed line MN in  Fig.\ref{fig_classification} represents weak external fields: $D < 0.1$  ($X = –2.3$). The dashed line AKB defines the boundaries of the area of relatively small energies for which $a<0.1$ ($Y=-2.3$). Accordingly, the AKN region $\it CD$ corresponds to the classical dipole approximation (Thomson backscattering, Eq.(\ref{quantum_dipole_spectrum}) for $u \ll 1$). The quantum dipole spectrum (Compton backscattering, Eq.(\ref{quantum_dipole_spectrum})) takes place in the AKM region $\it QD$. In the latter two regions, the radiation spectrum is characterized by high monochromaticity and displays a single peak terminating at the frequencies, $u_{mk}$,  given by the formula (\ref{u_mk}) for $k = 1$.

The condition for  CFA  ($D > 1$) in the quantum region ($a \ll 1+D^2$)  appears in Fig.\ref{fig_classification} as $2X > Y$, and the corresponding region extends to the right of the OP line. The radiation spectrum is determined by the quantum synchrotron formula (\ref{CFA_spectrum}) and has a shape determined by a single parameter $\chi \approx aD$. The region of the classical synchrotron approximation ($\chi < 0.1$; $ D \ge 1$) is determined by the inequality $X + Y < –2.3$ and corresponds to the {\it CS} sector (DBQ). In this case, the radiation spectrum appears as the classical synchrotron spectrum. 
By the same token, the $\it QS$  sector corresponds to the quantum synchrotron spectrum (QBOP). The validity of the CFA
 (\ref{CFA_spectrum}) depends also on the emitted photon energy and has the form \cite{mkh_2002} 
\begin{equation}
 a \ll (1 +D^2)u' .
\label{validity_of_CFA}
\end{equation}

The other regions shown in Fig.\ref{fig_classification}   require exact calculations taking into account the generation of higher harmonics.
The formulas of classical electrodynamics are valid in the NKBD region, whereas the exact quantum  expressions 
(\ref{photon_spectrum})-(\ref{Fourier_components_of_currents})  have to be used in the MKBOP region.
Detail explanation of the  inequality (\ref{validity_of_CFA})  is given in Supp.B. 

In the case of  LRS, all values of $a$ and $D$  in Fig \ref{fig_classification}  are available. The energy of an electron for LRS in Fig.\ref{fig_classification}  is $E=a(mc^2)^2/4\hbar \omega_0$. For a laser with $\hbar \omega_0 =1$ eV (this energy of the  laser photon will be used in numerical calculations below)  the origin of the coordinate system in Fig. \ref{fig_classification} corresponds to $E$=65.3 GeV.  

\begin{figure}
\centering
\includegraphics[bb=100 30 922 596, scale=0.36]{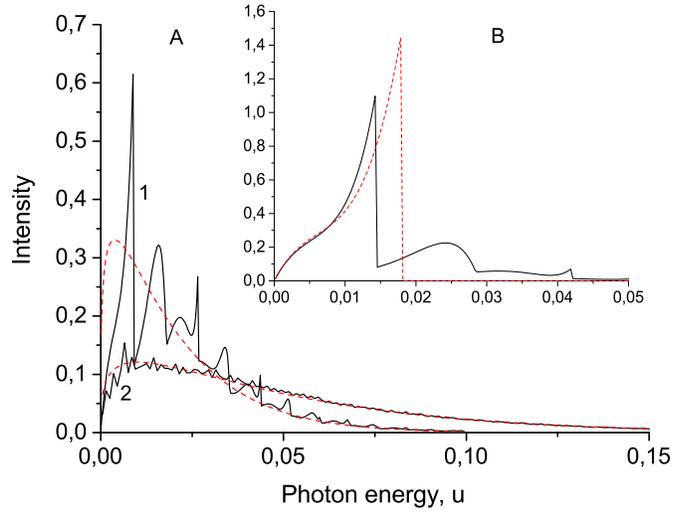}
\centering
\caption{
Radiation intensity spectra for $a=0.0183$ ($Y=-4$, centered triangles in Fig.\ref{fig_classification}). Solid lines are exact calculations according to (\ref{photon_spectrum}). A: $D$=1 ($X$=0, curve 1); $D$=2.718 ($X$=1, curve 2); dashed lines, spectra in the CFA  
 (\ref{CFA_spectrum}). B: $D$=0.5 ($X=-0.69$); dashed line, dipole approximation  (\ref{quantum_dipole_spectrum}). Beam energy for LRS is 1.2 GeV. 
}
\label{a_0_0183}
\end{figure}
\begin{figure}
\centering
\includegraphics[bb=50 30 922 596, scale=0.31]{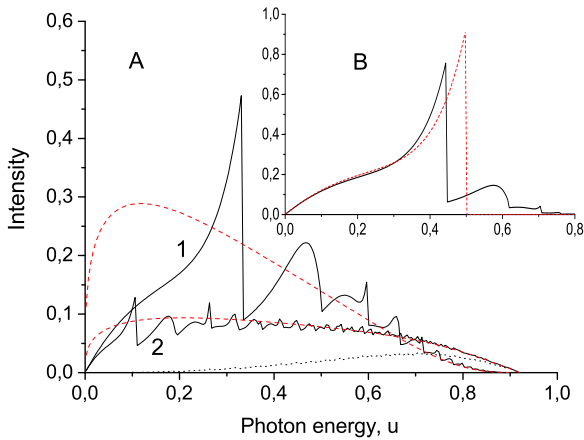}
\centering
\caption{
Radiation intensity spectra for $a=1$ ($Y=0$, centered circles in Fig.\ref{fig_classification}). Solid lines represent calculations according to Eq.(\ref{photon_spectrum}). A: $D$=1 ($X$=0, curve 1); $D$=2.718 ($X$=1, curve 2, dots show the spin contribution); dashed lines are spectra in the CFA  
(\ref{CFA_spectrum}). B: $D$=0.5 ($X=-0.69$); dashed line, dipole approximation  (\ref{quantum_dipole_spectrum}). Beam energy for LRS is 65.3 GeV. 
}
\label{a_1}
\end{figure}

In a crystal, only those values of $D$ and $a$ are possible that lie on the lines satisfying Eqs.(\ref{a_D_connection_planar}), (\ref{a_D_connection_axial}). Calculations are done for Si crystal ($Z=14$). 
The line 1 in Fig \ref{fig_classification}  represents the planar channeling of positrons  (\ref{a_D_connection_planar}) in Si (110) crystal,  $U_m=21$ eV, $d_p=1.92$  $\AA$ and $E_\perp =U_m$.  The  line 1 has  equation, $Y=X-4.48$. It  starts at positron energy $E\approx 31$ MeV and ends at $E\approx 5$ TeV. 
Axial channeling refers to the line 2 in Fig. \ref{fig_classification}. The calculation is for Si $\langle 110 \rangle$, $a_F=0.2$ $\AA$,  $d=3.84$ $\AA$.   The axial channeling line (\ref{a_D_connection_axial})  has the form, $Y=X-3.25$.
It starts at $E \approx$23 MeV end ends at $E \approx$3.7 TeV. 
Crosses on the lines 1 and 2 correspond  to the beam energy 300 GeV. Electron energies  of LRS at these points are 3.6 GeV for planar line 1 in Fig. \ref{fig_classification}  and 39 GeV for axial line 2. Beam energies for channeling lines in Fig. \ref{fig_classification}  can be calculated as   
\begin{eqnarray}
E & = &   \frac{a^2}{2U_m} \left( \frac{d_p}{4 \lambdabar_c} mc^2 \right)^2  \,,     \,\,\,\,\,\,\,\, \mbox{planar positrons},
\label{E_planar}
 \\ 
E & = &   \frac{a^2 d}{4Ze^2} \left( \frac{a_F}{2 \lambdabar_c} mc^2 \right)^2   \,, \,\,\,\,\,\,\,\, \mbox{axial electrons}.
\label{E_axial}
\end{eqnarray}

\begin{figure}
\centering
\includegraphics[bb=10 400 696 800, scale=0.43]{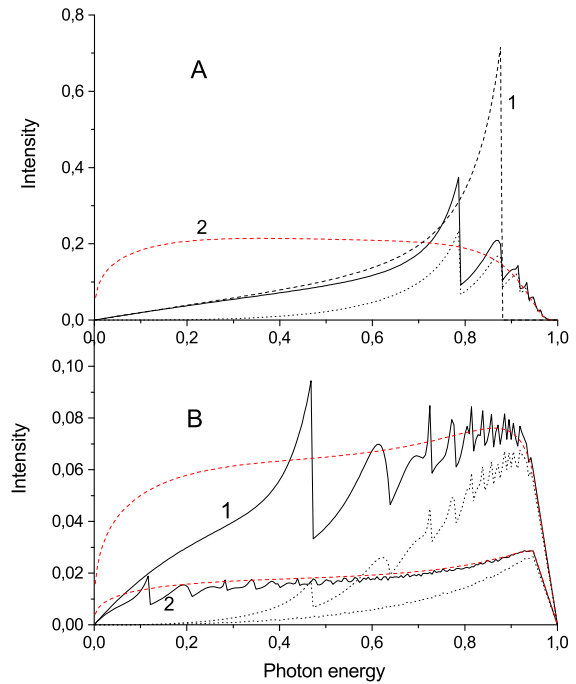}
\centering
\vspace{35mm}
\caption{
Radiation intensity spectra for $a=7.4$ ($Y=2$, centered squares in Fig.\ref{fig_classification}). Solid lines are exact calculations according to (\ref{photon_spectrum}). A: $D$=1 ($X$=0); dashed curve 1 is a dipole approximation (\ref{quantum_dipole_spectrum}); dashed curve 2 is CFA  
(\ref{CFA_spectrum}).  B:   $D$=2.718 ($X$=1, curve 1);  $D$=7.4 ($X$=2, curve 2);  dashed lines, spectra in the 
CFA  
(\ref{CFA_spectrum}). Dots show the spin contribution.  Beam energy for LRS is 483 GeV. 
}
\label{a_7_4}
\end{figure}

Calculations of the radiation intensity spectra (integrated over the azimuth angle $\varphi$ in (\ref{photon_spectrum}))  for points shown by the centered symbols in Fig.\ref{fig_classification} are presented in Figs. \ref{a_0_0183} -- \ref{a_7_4}.   Intensity is given in the units of $N_0$, i.e. the right hand sides of Eqs.(\ref{photon_spectrum}),  (\ref{quantum_dipole_spectrum}),  (\ref{CFA_spectrum}),  multiplied by $u=\hbar \omega /E$ are shown as functions of $u$. Fig.\ref{a_0_0183} represents the classical radiation spectra, $a \ll 1$, for the points shown by triangles in Fig. \ref{fig_classification}. The LRS electron beam energy is 1.2 GeV. In this case CFA adequately describes the spectrum already for $D>2$.   

Quantum effects of emitted photon recoil and spin become clearly expressed at $a=1$, as is seen in Fig.\ref{a_1} (circles  in Fig.\ref{fig_classification}). The LRS energy in this case is 65.3 GeV.   For $ D = 1$  CFA is applicable only for  hard photons in accordance with condition (\ref{validity_of_CFA}), but for $D = 2.78$ CFA describes the spectrum well, except for the relatively low-frequency region, where individual harmonics are clearly pronounced.

Intensity spectra for very high energies (483 GeV for LRS) are shown in Fig. \ref{a_7_4}  ($a=7.4$, $Y=2$, square characters in the Fig.\ref{fig_classification}). The spin contribution dominates at high energy photon region, $u>0.6$. CFA describes the radiation spectrum at relatively large values of $D>7$. As it follows from Fig. \ref{a_7_4}, even for $D=7.4$ there is a pronounced peak in the soft part of the spectrum. 

It is well known, that the validity of the CFA in channeling becomes better while the electron energy increases, since $D \sim \gamma^{1/2}$. 
There is a reverse trend in the case of LRS. For fixed laser field intensity, $D$,  the justness of CFA for given photon energy interval  worsens  with an increase in electron energy.  This is clear from Figs. \ref{a_0_0183} -- \ref{a_7_4} for $D=2.718$ (symbols in  Fig. \ref{fig_classification} for $X=1$). At low energies (curve 2 in Fig. \ref{a_0_0183}) CFA describes the whole radiation spectrum. At higher energy corresponding to curve 2 in Fig. \ref{a_1} individual harmonics are clearly visible in the soft part of the spectrum, while at very high energies (curve 1 in Fig. \ref{a_7_4}, B), CFA is valid only in the hard photon region, $u>0.9$.       

 There is a significant difference between the dependence of  parameters $D$ and $a$ on the transverse energy for channeling (presented in Table 1) and quasi-channeling (see Supp.D). The parameter $ D$ for quasi-channeling decreases with increasing the transverse energy as $\sim E_\perp^{-1/2}$. The applicability of the CFA becomes less acceptable as the transverse energy increases. This agrees with the results of Refs. \cite{mkh_Nitta_2002, mkh_Nitta_2004,  mkh_Nitta_2005}, in which the field non-uniformity parameter, $\nu$, coincides with the quasi-channeling parameter $ D$  (see Eq.(D3) in  Supp.D) up to a factor of the order of unity (Eq.(12) in \cite{mkh_Nitta_2002}).

\vspace{-5mm}


\section{Summary and concluding remarks}

Radiation spectrum of ultrarelativistic electrons (positrons) in the fields of lasers and oriented crystals can be expressed in terms of two invariant parameters, one of which, $a$,  characterizes the significance of quantum effects in radiation and depends on the electron (positron) energy, while the non-dipole  parameter $D$ determines the influence of the external field. In the particular case of the linear polarized laser wave and planar positron channeling the radiation spectra are  described by the same formula (\ref{photon_spectrum}), where parameters $a$ and $D$  are determined by  expressions shown in the Table 1.  The crucial difference between LRS  and CR is that in the former case  parameters $a$ and $D$ are independent, while in channeling they are linearly related to each other. This leads to a strong limitation on the range of possible values of the parameters $a$ and $D$ for channeling, since they lie on a line in the space of $X=\ln D$ and $Y=\ln a$, as shown in the diagram on Fig.\ref{fig_classification} (the $D-a$ line). For example, the  quantum dipole radiation spectrum  with a single peak in the hard photon region, $\hbar \omega \sim E$,  is not possible in channeling, while for LRS it corresponds to the Compton back scattering in the	 weak field, $D \ll 1$, $a \ge 1$.   

The $D$ parameter for LRS does not depend on particle's energy, while for CR it increases as $\sim \gamma^{1/2}$.  
The consequence of this is that the applicability of the CFA for channeled particles becomes better as their energy increases. For LRS, conversely, the condition for CFA applicability is violated as the energy grows for fixed $D$.

In contrast to channeled particles, the non-dipole parameter $D$ for quasi-channeled particles decreases with increasing the transverse energy. 
Accordingly, the $D-a$  line of quasi-channeled particles, Eq.(D4) in Supp.D, 
 lies the higher, the greater their transverse energy, and the applicability of the CFA deteriorates. 
The influence of the spatial non-uniformity of the field of atomic chains (planes) on radiation becomes stronger. For example, a strong suppression of radiation for TeV quasi-channeled electrons in about the whole spectrum compared with that for CFA may take place \cite{mkh_Nitta_2002}. Thus, the OC makes it possible to study QED processes in strong non-uniform external fields.





\section*{Supplementary materials}

\subsection*{A. Exact classical equations of motion for LRS}  

 It is reasonable to compare Eqs.(\ref{pl_las_eq_motion}) with exact solutions. In the case considered the exact classical equations of motion for LRS  have the form  
$$
x  =  x_m \, \sin \delta \, ,
$$
\vspace{-7mm}
$$
z  =  \beta_0 ct - \frac{x_m^2 \Omega_0}{8c} \, \sin 2\delta,
\eqno(\mbox{A}1)
$$
\vspace{-7mm}
$$
\Omega_0 t  =   \delta + \frac{x_m^2 \Omega_0^2}{8c^2 (1+\beta_0)} \sin 2\delta \, , 
$$
$$
\gamma (\delta)  =  \gamma_0 \sqrt {1+\nu_0^2} \left[ 1+ \frac{x_m^2 \Omega_0^2}{4c^2 (1+\beta_0)} \cos 2\delta
\right],
\eqno(\mbox{A}2)  
$$
where  the quantities $\Omega_0$ and $x_m$ are given in the Table 1. 

Eqs. (A1)
and (A2)
satisfy the Lorentz equations of motion, 
$mcdv^\mu /d\tau = e F^{\mu\nu} v_\nu$, where $v^\mu$ is a 4-velocity, $\tau$ is a proper time. Space components of  this equations give 
(A1), while the time component leads to the  
Eq.(A2).
 It follows from (A1) and (A2)  
that, 
$\gamma^2 (\delta) \beta_x^2 (\delta) = 2\nu_0^2 \cos^2 \delta$.  Therefore, the average over the phase gives for LRS, 
$D \equiv \langle \beta_\perp (\delta)^2 \gamma (\delta)^2 \rangle^{1/2} = \nu_0$. The average Lorentz-factor, according to Eq.(A2),
is $ \langle \gamma (\delta) \rangle =  \gamma_0 \sqrt {1+\nu_0^2}$. This value should be taken as the initial Lorentz-factor of the electron before the interaction with the external field, i.e. in infinity.   

At high energies, if conditions (\ref{inequalities}) take place, the term in the third line of 
Eq.(A1),
$(x_m \Omega_0/c)^2 \sim (D/\gamma)^2 \ll 1$,   is negligibly small, such that we can replace the arguments of sine in the first two lines by, $\delta \approx \Omega_0 t$. We arrive then to Eqs.(\ref{pl_las_eq_motion}). Note, that in this limit the oscillating term in Eq.(A2)
is also negligible, and the Lorentz-factor is time independent, equal to its initial value,  $ \gamma =  \gamma_0 \sqrt {1+D^2}$.

\subsection*{B. Quasi-periodic motion and formation time of radiation }  

{\em Radiation during quasi-periodic motion} 

\vspace{2mm}

Within the frame of the quasiclassical BK method, which applicability is consistent with conditions (\ref{inequalities}),  the spectral and angular
distribution of energy emitted by a relativistic electron moving along an arbitrary
trajectory is \cite{Baier1998}
$$
I_{\omega \bf n } \equiv \frac{d^2 I}{d\omega d \Omega}  =  \frac{e^2 \omega^2}
{8\pi^2 c} \frac{E^2+E'^2}{E'^2}  \times
$$
\vspace{-5mm}
$$
\eqno(\mbox{B}1)  
$$
\vspace{-5mm}
 $$ 
 \times  \left ( \mid {\bf n} \times
[{\bf n} \times {\bf j}_{\omega}] \mid ^2 + \frac{\hbar^2 \omega^2 \gamma^{-2}}
{E^2+E'^2}J \right )  ,      \nonumber
$$
where $E'=E-\hbar \omega$  is the final electron energy,  ${\bf n} = ( \sin \theta \cos \varphi, \sin \theta \sin \varphi, 
\cos \theta)$  is a unit vector in the direction of radiation,
$$
{\bf j}_\omega = \int_{-\infty}^{+\infty}
\mbox{\boldmath $\beta$}
(t) \exp \Bigl(i \omega' (t-{\bf rn}/c) \Bigr) dt , 
$$ 
\vspace{-8mm}
$$
\eqno(\mbox{B}2)  
$$
\vspace{-8mm}
$$
J  =  \left|
\int_{-\infty}^{+\infty} \exp
\Bigl( i \omega' (t-{\bf rn}/c) \Bigr) dt \right|^2  ,
$$
where $\omega'=\omega E/(E-\hbar \omega)$. 
In practice, semiclassical formulas (B1), (B2) are usually used in the small-angle approximation, $\theta \ll 1$,  $\theta_e  \ll 1$,  since, strictly speaking, they are valid only in this case. This follows especially clearly from the derivation of semiclassical formulas in \cite{Lindhard_91}.

For any quasiperiodic motion, ${\bf r}_\perp (t)={\bf r}_\perp (t+T)$, given, for example, by Eqs. (\ref{pl_las_eq_motion}), $\Omega_0=2\pi /T$, general expressions (B1) and (B2) 
 give the following result for the power spectrum ($\theta \ll 1$, $\beta_\perp \ll 1$)  
$$
P_{\omega \bf n}  \equiv   \frac{d^3 I}{d\omega d\Omega dt}  = 
\frac{e^2 \omega^2}{4 \pi c} \frac{(E^2+E'^2)}{E'^2} \times 
\eqno(\mbox{B}3)
$$
$$
 \times \sum^{\infty}_{k=1}
Q(k,{\bf n},\omega) 
 \delta
  \left \{ \frac{\omega'}{2}(\theta^2+\gamma^{-2}
  + \langle \mbox{\boldmath $\beta$}^2_\perp \rangle ) -k \Omega_0  \right \},
$$
$$
Q(k,{\bf n},\omega)= \, | {\bf j}_{\perp k} |^2  -  \gamma^{-2} | j_{z k} |^2  -
$$
$$
 -  Re ( j_{2k}j_{zk}^\ast ) + \frac{\hbar^2 \omega^2 \gamma^{-2}}
{E^2+E'^2} \, | j_{zk} |^2 ,
$$
where
\[{\bf j}_{\perp k}=\frac{1}{T}\int_0^T \mbox{\boldmath $\beta$}_\perp (t)
\exp (i\Delta_k) dt, \]
\vspace{-7mm}
$$
j_{z k}=\frac{1}{T}\int_0^T
\exp (i\Delta_k) dt,
\eqno(\mbox{B}4)  
$$
\vspace{-7mm}
\[j_{2 k}=\frac{1}{T}\int_0^T \mbox{\boldmath $\beta$}^2_\perp (t)
\exp (i\Delta_k) dt ,\]
\vspace{-7mm}
$$
\Delta_k=\,\Omega_0 kt-\frac{\omega'}{c}\;\theta \: {\bf n}_\perp {\bf r}_\perp -
\frac{\omega'}{c}\;\delta z(t) ,
\eqno(\mbox{B}5)
$$  
where ${\bf n}_\perp = (\cos \varphi, \sin \varphi)$, $\varphi$ is the azimuthal
angle of radiation, $\theta$ is the polar angle of radiation with respect to the atomic axes (planes),
$\sin \theta \approx \theta$,\,\, $\cos \theta \approx \, 1-\theta^2/2$.
The longitudinal oscillations in (B5) are
$$
\delta z(t) = \frac{c}{2}\int^{t}_{0}[\langle  \mbox{\boldmath $\beta$}^2_\perp \rangle-
\mbox{\boldmath $\beta$}^2_\perp (t)]dt . 
\eqno(\mbox{B}6)
$$ 

For trajectories given by Eqs. (\ref{pl_las_eq_motion})  expressions  (B3)--(B6) lead to the results 
(\ref{photon_spectrum}) -- (\ref{Fourier_components_of_currents}) after the integration over the radiation angle $\theta$. Formula for $u_{mk}$ in Eq. (\ref{u_mk}) comes from the argument of the $\delta$-function in (B3). Eqs. (B3)--(B6) describe both planar and axial channeling with closed transverse trajectories. General formulas for axial channeling with allowance for precession of the transverse orbits have been obtained in Ref. \cite{mkh_1984}.

\vspace{2mm}

{\em Constant field approximation}

\vspace{2mm}

Under conditions when a large number of harmonics, $k\gg 1$, contribute to the radiation, but the  spectrum is not yet of the synchrotron type, like in Eq.(\ref{CFA_spectrum}),
the summation in (B3) can be replaced by the integration over $k$
$$
\sum^{\infty}_{k=1} (...) \rightarrow \int_0^\infty (...)dk . 
\eqno(\mbox{B}7)
$$

The integration over $k$ in Eq.(B3) gives
$$
P_{\omega \bf n} = 
\frac{e^2 \omega^2}{4 \pi c \Omega_0} \frac{(E^2+E'^2)}{E'^2} \times 
\eqno(\mbox{B}8)
$$
\vspace{-5mm}
$$
\times Q \left \{ k=\frac{\omega'}{2\Omega_0}(\theta^2+\gamma^{-2}
  + \langle \mbox{\boldmath $\beta$}^2_\perp \rangle ),  {\bf n}, \omega  \right \},
$$
where $k$ can be expressed in terms of the invariant quantities
$$
k=\frac{u'}{a}\left( 1+ \eta^2+D^2   \right),
\eqno(\mbox{B}9)
$$
with $\eta=\theta \gamma$ (note that the value of $k$ in (B8) coincides with  corresponding formula in \cite{Artru2015}). For radiation in the ``forward'' direction, $\theta=0$, the condition, $k\gg 1$, gives  inequality f (\ref{validity_of_CFA})
$$
\frac{u'}{a}\left( 1+D^2   \right) \gg 1. 
\eqno(\mbox{B}10)
$$

Formula (B8) is more accurate than the CFA formula (\ref{CFA_spectrum}), especially in the soft frequency range, where CFA  gives an artificial divergence for the photon spectrum $\sim \omega^{-2/3}$.

\vspace{2mm}

{\em Formation time of radiation}

\vspace{2mm}



 Formation time,  $t_C$, which we estimate up to the factor of the order of unity, corresponds to the phase, $\Delta \approx 2\pi$. With these definitions we obtain 
$$
t_C = T \frac {a}{u'(1+D^2)},
\eqno(\mbox{B}11)
$$ 
where $T=2\pi /\Omega_0$ is the period of the transverse motion. The CFA is valid if $t_C \ll T$, which condition is equivalent to (B10). In the dipole approximation, when $D \ll 1$, (B11) gives well known expression for the formation time, $t_C = 4\pi \gamma \gamma' /\omega$, where $\gamma'$ is the Lorentz-factor after the photon emission. 

We did not use expressions for the formation time of radiation in a constant external field, since we did not use expansions of the phase (B5)  in powers of the time variable. 
However, it is interesting to compare the above expressions with the case of  CFA  in terms of the combination 
$D^2u'/a$. The conventionally used expression (up to a factor of the order of unity) for radiation formation time in the constant external field is
 \cite{Pedersen1987,Artru1988} 
$$
t_C \approx  \left( \frac{EE'}{F^2c^2\omega}    \right)^{1/3}  \approx 
T \left( \frac{a}{D^2u'}    \right)^{1/3},
\eqno(\mbox{B}12)
$$
where $F$ is the force. More precise formula for constant field formation time, derived by means of the complex time method, is given in Ref.\cite{mkh2010}. It has been shown, that the radiation formation time (length)
for low energy photons decreases upon an increase in the radiation frequency in accordance with the familiar law $t_C \sim \omega^{-1/3}$ (B12), while for higher frequencies, this dependence changes to 
$t_C \sim \omega^{-1/2}$. 

Clear conclusions about the difference between (B11) and (B12) are given in Ref.\cite{Artru2015}.  Eq.(B11) (for $\theta=0$), equivalent to,  $t_C=T/k_m$, gives a formation time for one period of motion, $ k_m$  is given by (B9) with $\theta=0$. Eq.(B12) is the formation time when $  \mbox{\boldmath $\beta$} $ is parallel to ${\bf n}$ and concerns only part of the period of motion. When  $  \mbox{\boldmath $\beta$} $ is not parallel to ${\bf n}$, one may use Eq.(10) from \cite{Artru2015}:
$$
t_C \approx \min \{ t_C' , t_C''  \}
$$
\vspace{-8mm}
$$
\eqno(\mbox{B}13)
$$
\vspace{-6mm}
$$
t_C'=\frac {4\pi / \omega'}{\gamma^{-2} + (  {\bf n }  -  \mbox{\boldmath $\beta$})^2 }, \, \, \, \, \, \, 
t_C''= 2\pi \left(   \frac{24R^2}{c^2 \omega' }    \right)^{1/3} ,
$$
where $R=E/\!\! \!\! \mid \!\! \!U ({\bf r}_\perp) \!\!\! \mid $ is the local  bending radius, $\omega'=\gamma \omega /\gamma'$,  $t_C''$ corresponds  to (B12) but with the ``true'' coefficient.  When $ \mid \! {\bf n }  -  \mbox{\boldmath $\beta$} \! \mid $ is large , $t_C=t_C'$. For example, for $ \mid \! {\bf n }  -  \mbox{\boldmath $\beta$} \! \mid \sim D/\gamma$ one has $t_C=t_C' \sim T/k_m$.

\subsection*{C. Comment to axial channeling}  

 The connection between invariant parameters $a$ and $D$ for planar positron channeling, given by Eq.(\ref{a_D_connection_planar}),  is precise, whereas for axial channeling, Eq.(\ref{a_D_connection_axial}), it is approximate. Approximation for axial case is based on the relation, 
$$
\Omega_0 \approx c\theta_L/a_F .
\eqno(\mbox{C}1)
$$. 
Exact formula for the frequency of the transverse elliptic motion in the Coulomb-like atomic string potential, $U(r)=-\alpha /r$, is \cite{Beloshitsky1982} 
$$
\Omega_0 = \frac{ 2 \mid \! \! E_\perp \! \! \mid c}{\alpha}\left(  \frac{ 2 \mid \! \! E_\perp \! \! \mid }{E} \right)^{1/2}, 
\eqno(\mbox{C}2)  
$$
where $\alpha = A Ze^2 a_F/d$, with $A \approx 3/2$ \cite{Beloshitsky1982, Lindhard1965}. The transverse energy for channeling electrons is negative, $E_\perp < 0$. For  typical values, 
$\mid \! \! E_\perp \! \! \mid \approx  Ze^2 /d$, and up to the coefficient on the order of unity, (C2) leads to (C1).

\subsection*{D. Quasi-channeling}  

The transverse trajectories of the above-barrier particles are unbound, $E_\perp > U_m$. For LRS  this kind of trajectories does not exist with except of the scattering of the incoming particle beam by the intense laser beam with $\nu_0 > \gamma$. 
In planar quasi-channeling, the above-barrier particles cross atomic planes with a period, $T_q$, equal to the interplanar travel time. 
The definition of the invariant $D$ in this case should take into account that the average transverse velocity is not zero: 
$D=\langle \Delta \beta_\perp^2 \rangle^{1/2} \gamma$, where  
$\langle \Delta \beta_\perp^2 \rangle = \langle \beta_\perp^2 \rangle - \langle \beta_\perp \rangle^2 $. 
We define also the invariant $a$ for quasi-channeled particles as:  $a=2\hbar \gamma \Omega_q /(mc^2)$, where
$\Omega_q= 2\pi/T_q$.  

The mean square variation of the transverse velocity can be calculated analytically for some realistic model potentials. For planar positron quasi-channeling the result is
$$
\langle \Delta \beta_\perp^2 \rangle=\frac{U_m}{E}f(y),
\eqno(\mbox{D}1)
$$
\vspace{-4mm}
$$
f(y)=\frac{1}{s} \sqrt{ \frac{1}{y^2}-1 } +\frac{1}{y^2} - \frac{2}{s^2} ,   
\eqno(\mbox{D}1a)
$$
where $y=(U_m/E_\perp)^{1/2}\le 1$, $s=\arcsin y$.
For $y\ll 1$, $f(y)\approx (2/45)y^2$. With $y=0.9$, this formula gives a two times lower result compared to exact Eq.(D1a), while for $y=0.5$ and $y=0.2$ its accuracy is 13\% and 2\%, correspondingly.  
Note, that when passing through the border from channeling to quasi-channeling, the value of $D$ (or $\langle \Delta \beta_\perp^2 \rangle$  in (D1)) suffers a break, namely, it decreases by 0.44 times compared to channeling. This comes from the contribution of the average transverse velocity (the last term in (D1a)), which is zero during channeling, but immediately takes on a noticeable value when the positron (electron) leaves the channel.

We shall restrict ourselves with the case of large transverse energies, $E_\perp \gg U_m$.  In this limit the result is: 
$\langle \Delta \beta_\perp^2 \rangle \approx \xi U_m^2/(E_\perp E)$, where for planar  quasi-channeling in the parabolic potential $\xi = 2/45$.    This formula is valid up to the terms $\sim (U_m/E_\perp)^2$. 
Calculations  for planar electron quasi-channeling in the potential, $\sim \cosh^{-2} (kx)$, give  $\xi =0.028$ (in Si (110), $U_m=$22.9 eV, $1/k=$0.303 $\AA$).

For invariant parameters $a$ and $D$ we obtain  
$$
a=4\pi \frac{ \lambdabar_c}{d_p}\left( \frac{2 E_\perp \gamma}{mc^2}  \right)^{1/2},
\eqno(\mbox{D}2)
$$

\vspace{-4mm}

$$
D=U_m \left( \frac{\xi \gamma}{mc^2  E_\perp}  \right)^{1/2} .
\eqno(\mbox{D}3)
$$

The corresponding line on the   $D \div a$ diagram is
$$
a=4\pi \frac{ \lambdabar_c}{d_p} \frac{ E_\perp }{ U_m}   \left( \frac{2 }{\xi}  \right)^{1/2} D .
\eqno(\mbox{D}4)
$$


The positron energy on the line (D4) is
$$
E  =    \frac{a^2}{2E_\perp} \left( \frac{d_p}{4 \pi \lambdabar_c} mc^2 \right)^2  .
\eqno(\mbox{D}5)
$$

Formulas (D1) -- (D5)
demonstrate a significant difference between the dependence of  parameters $D$ and $a$ on the transverse energy for channeling (presented in the Table 1) and quasi-channeling. 
The parameter $ D$ for quasi-channeling decreases with increasing the transverse energy as $\sim E_\perp^{-1/2}$, which causes the radiation to become more dipole and the transverse trajectories to become more rectilinear, so that the transverse velocity tends to $\beta_\perp \rightarrow (2E_\perp/E)^{1/2}$.
The applicability of the CFA becomes less acceptable as the transverse energy increases. 

For 300 GeV positrons in Si (110) and $E_\perp/U_m =10$ (it corresponds to the angles $\sim 3\theta_L$ with respect to the atomic plane):  $D=$0.33 and $a=$0.55, which values belong to the region of quantum dipole spectrum in the diagram on Fig.\ref{fig_classification}. The $D-a$ line  for quasi-channeling (D4)
lies above the line for channeled particles. Its height increases with the transverse energy
(not shown in Fig.\ref{fig_classification}). 
For example, for   $E_\perp/U_m =10$ the quasi-channeling line (D4)
is parallel to lines 1 and 2 on the diagram 
in Fig.\ref{fig_classification} and goes through the point $Y=0.53$ and $X=0$ ($D=1$). 
The dipole approximation for such positrons  is violated at TeV energies, ($D=1$ for $E=2.8$ TeV when $a=1.7$).

We can say that CFA  is valid if the angles of entry of particles into an oriented crystal do not exceed the critical Lindhard angle, $\theta_L$. Since in real crystals the transverse energy tends to increase due to multiple scattering,  the particles continuously move to higher $D-a$  lines as they penetrate through the target.


\begin{thebibliography}{33}

\addcontentsline{toc}{section}{References}

\bibitem{Beloshitsky1982}
V.V. Beloshitsky, F.F. Komarov,  Phys. Rep. 93 (1982) 117-197.
\bibitem{Ulrik2005}
U.I. Uggerh\o j, Rev. Mod. Phys.  77 (2005) 1131-1171.
\bibitem{Ulrik_PLB_2017}
A. Di Piazza, T.N. Wistisen, U.I. Uggerh\o j,  Phys. Lett. B 765 (2017) 1-5.
\bibitem{Ulrik_NC_2018}
T.N. Wistisen, A. Di Piazza, H.V. Knudsen, U.I. Uggerh\o j,  Nat. Commun. 9 (2018) 795.
\bibitem{mkh_2019}
M.Kh.Khokonov, Phys. Lett. B 791 (2019) 281-286.
%
\bibitem{Wistisen_2019}
T.N. Wistisen, A. Di Piazza, C.F. Nielsen, A.H. Sorensen, U.I. Uggerh\o j, Phys. Rev. Res. 1 (2019) 033014.
%
\bibitem{Wistisen_2020}
Ch.F. Nielsen, J.B. Justesen, A.H. Sorensen, U.I. Uggerh\o j, R. Holtzapple, Phys. Rev. D 102 (2020) 052004. 


\bibitem{uggerhoj_2020}
A. Di Piazza, T.N. Wistisen, M. Tamburini, U.I. Uggerh\o j, Phys. Rev. Lett. 124 (2020) 044801. 

\bibitem{uggerhoj_2021}
F.C. Salgado, N. Cavanagh, M. Tamburini, et al., New J. Phys. 24 (2021) 015002. 

\bibitem{trident_2010}
J. Esberg, K. Kirsebom, H. Knudsen, et al., Phys. Rev. D 82 (2010) 072002. 

\bibitem{trident_2022}
Ch.F. Nielsen, R. Holtzapple, M.M. Lund, et al., arXiv:2211.02390 [hep-ex]. 




\bibitem{Keitel_RMP2012}
A. Di Piazza, C. Muller,  K.Z. Hatsagortsyan,  Ch.H. Keitel, Rev. Mod. Phys. 84 (2012) 1177-1228. 


\bibitem{LUXE_2021}
H. Abramowicz, et al. Conceptual design report for the LUXE experiment,
Europ. Phys. J. Special Topics,  230, (2021) 2445-2560.

\bibitem{mkh_2002}
A.Kh.Khokonov, M.Kh.Khokonov, A.A.Kizdermishov, Tech. Phys. 47 (2002) 1413-1419. 

\bibitem{King_2020}
B. King, S. Tang, Phys. Rev. A 102 (2020) 022809.

\bibitem{Ritus1985}
V.I. Ritus, J. Russ. Laser Res. 6  (1985) 497-617; translated from: Tr. Fiz. Inst. Akad. Nauk SSSR 111 (1979) 5-151.

\bibitem{Gonoskov_RMP_2022}
A. Gonoskov, T.G. Blackburn, M. Marklund, S.S. Bulanov, Rev. Mod. Phys. 94 (2022) 045001.  

\bibitem{Piazza_PRD_2019}
T. Podszus, A. Di Piazza, Phys. Rev. D 99 (2019) 076004.

\bibitem{Ilderton_PRD_2019}
A. Ilderton, Phys. Rev. D 99 (2019) 085002.

\bibitem{Mironov_PRD_2019}
A. Mironov, S. Meuren,  A. Fedotov, Phys. Rev. D 102 (2020) 053005.

\bibitem{Kostykov_Sci_2019}
C. Baumann, E.N. Nerush, A. Pukhov, I.Y. Kostyukov, Sci. Rep. 9 (2019) 9407.

\bibitem{Keitel_PRL_2010}
H. Hu, C. Muller,  C.H. Keitel, Phys. Rev. Lett. 105 (2010) 080401.

\bibitem{Ilderton_PRL_2019}
A. Ilderton, Phys. Rev. Lett. 106 (2011) 020404.

%
\bibitem{Carrigan1998}
M.Kh.Khokonov, R.A. Carrigan Jr., Nucl. Instrum. Methods Phys. Res. B 145 (1998) 133-141.

\bibitem{Jackson}
J.D. Jackson, Classical Electrodynamics, John Wiley \& Sons. Inc., New York, 1999.
\bibitem{mkh_NIMB1998}
A.Kh. Khokonov, M.Kh. Khokonov, R.M. Keshev, Nucl. Instrum. Methods Phys. Res. B  145 (1998) 54-59.

\bibitem{Lindhard1965}
J. Lindhard, Kgl. Dan. Vid. Selsk. Mat. Fys. Medd. 34(14) (1965).

\bibitem{Beloshitsky_Kumakhov_1978}
V.V. Beloshitsky, M.A. Kumakhov, Sov. Phys. JETP 47 (1978) 652-658.

\bibitem{Zhevago1978}
N.K. Zhevago, Sov. Phys. JETP 48 (1978) 701-707.  

\bibitem{Kumakhov_Trikalinos_1980}
M.A. Kumakhov, Kh.G. Trikalinos, Sov. Phys. JETP 51 (1980) 815-821.  

\bibitem{Wistisen_oscillator_2018}
T.N. Wistisen, A. Di Piazza, Phys. Rev. A 98 (2018) 022131. 

\bibitem{Wistisen_planar_2019}
T.N. Wistisen and A. Di Piazza, Phys. Rev. D 99 (2019) 116010.

\bibitem{Baier1998}
V.N. Baier, V.M. Katkov, V.M. Strakhovenko,  Electromagnetic Processes at High Energies in 
Oriented Single Crystals,  World Scientific Pub Co Inc, 1998. 

\bibitem{Lindhard_91}
J. Lindhard, Phys. Rev. A 43 (1991) 6032-6037.

\bibitem{Artru2019}
X. Artru, Phys. Rev. Accel. Beams 22 (2019) 050705.

\bibitem{Artru2015}
X. Artru,  Nucl. Instrum. Methods Phys. Res. B 335 (2015) 11-16. 

\bibitem{Harvey2009}
C. Harvey, T. Heinzl, A. Ilderton, Phys. Rev. A 79 (2009) 063407.

\bibitem{Ritus1964a}
A.I. Nikishov, V.I. Ritus, Sov. Phys. JETP 19 (1964) 529-541. 

\bibitem{Baier1980}
V.N. Baier, V.M. Katkov, V.M. Strakhovenko, Zh. Eksp. Teor. Fiz. 80  (1981)  1348-1360.

\bibitem{Planar_positrons_1981}
V.A. Bazylev, V.V. Beloshitsky, V.I. Glebov et al., Sov. Phys. JETP 53 (1981) 306-316. 

\bibitem{Walker_1998}
R. Walker. Insertion devices: undulators and wigglers. In:
CERN accelerator school. Synchrotron radiation and free electron lasers. Geneva,  p. 129, 1998. 

\bibitem{Klepikov1954}
N.G. Klepikov, Zh. Eksp. Teor. Fiz. 26 (1954) 19. 

\bibitem{Khokonov1997}
M.Kh. Khokonov. Phys. Scripta 55  (1997) 513-519.

\bibitem{Khokonov2004}
M.Kh. Khokonov, Sov. Phys. JETP 99 (2004) 690-707.

\bibitem{Khokonov2005}
A.Kh. Khokonov, M.Kh. Khokonov, Tech. Phys. Lett. 31 (2005) 154-156.

\bibitem{mkh_Nitta_2002}
M.Kh. Khokonov, H. Nitta,  Phys. Rev. Lett. 89 (2002) 094801.

\bibitem{mkh_Nitta_2004}
H. Nitta, M.Kh. Khokonov, Y. Nagata, S. Onuki, Phys. Rev. Lett.  93 (2004) 180407. 


\bibitem{mkh_Nitta_2005}
Y. Nagata, H. Nitta, M.Kh. Khokonov, Nucl. Instrum. Methods Phys. Res. B 234 (2005) 159-167. 

\bibitem{mkh_1984}
N.K. Zhevago, M. Kh. Khokonov, Sov. Phys. JETP 60 (1984) 33-42.  

\bibitem{Pedersen1987}
 O. Pedersen, J.U. Andersen, and E. Bonderup, in Relativistic Channeling, Ed. by R.A. Carrigan and 
J.A. Ellison (Plenum, New York, 1987), NATO ASI, Ser. B, pp. 207-226.

\bibitem{Artru1988}
X. Artru, Phys. Lett. A 128  (1988) 302.

\bibitem{mkh2010}
M.Kh.Khokonov, I.Z.Bekulova,  Technical Physics, 55 (2010) 728-731.


\end{thebibliography}
\end{document}